\documentclass{article}
\pdfoutput=1
\usepackage{microtype}
\usepackage{graphicx}
\usepackage{subfigure}

\usepackage{booktabs} 
\usepackage[table,xcdraw]{xcolor}
\usepackage{hyperref}

\usepackage[accepted]{icml2019}

\usepackage{amsmath}

\usepackage{xcolor}

\icmltitlerunning{Coupling Oceanic Observation Systems to Study Mesoscale Ocean Dynamics}

\begin{document}
\twocolumn[
\icmltitle{Coupling Oceanic Observation Systems to Study Mesoscale Ocean Dynamics}



\icmlsetsymbol{equal}{*}

\begin{icmlauthorlist}
\icmlauthor{Gautier Cosne}{IMT}
\icmlauthor{Guillaume Maze}{equal,maze}
\icmlauthor{Pierre Tandeo}{equal,tandeo}

\end{icmlauthorlist}
\icmlaffiliation{IMT}{IMT Atlantique, Technop\^ole Brest-Iroise\\ Brest, France\\}
\icmlaffiliation{tandeo}{IMT Atlantique, Lab-STICC, UMR CNRS 6285, F-29238, France}
\icmlaffiliation{maze}{Ifremer, University of Brest, CNRS, IRD, Laboratoire d'Océanographie Physique et Spatiale (LOPS), IUEM, Brest, France}

\icmlcorrespondingauthor{Gautier Cosne}{gautier.cosne@imt-atlantique.net}

\icmlkeywords{Ocean modelling, Eddy-resolving, Operational Oceanography, North Atlantic, Unsupervised Classification, Latent Class Regression}

\vskip 0.3in
]



\printAffiliationsAndNotice{\icmlEqualContribution} 

\begin{abstract}

Understanding local currents in the North Atlantic region of the ocean is a key part of modelling heat transfer and global climate patterns. Satellites provide a surface signature of the temperature of the ocean with a high horizontal resolution while in situ autonomous probes supply high vertical resolution, but horizontally sparse, knowledge of the ocean interior thermal structure. The objective of this paper is to develop a methodology to combine these complementary ocean observing systems measurements to obtain a three-dimensional time series of ocean temperatures with high horizontal and vertical resolution. Within an observation-driven framework, we investigate the extent to which mesoscale ocean dynamics in the North Atlantic region may be decomposed into a mixture of dynamical modes, characterized by different local regressions between Sea Surface Temperature (SST), Sea Level Anomalies (SLA) and Vertical Temperature fields. Ultimately we propose a Latent-class regression method to improve prediction of vertical ocean temperature.
\end{abstract}

\section{Introduction}
\label{submission}

The ocean is a turbulent and chaotic dynamic system, very complex to observe, understand and to predict. This is especially true off the northeast coast of the United States where the Gulf Stream shrinks to intense fronts, meanders, and numerous eddies.  
For instance, \citet{Eddies}~shows that oceanic transports of heat of mesoscale eddies, which covers scales from 50 to 200 km$^2$, is comparable in magnitude to that of the large-scale wind and thermohaline-driven circulation. Understanding their thermal structure is key to follow the diffusion and transport of heat in the ocean. 

However, none of the existing ocean observing systems are able to provide an estimate of 3D thermal field with enough resolution to study individual eddies of mesoscale.
Satellite observations have a high local frequency and high horizontal resolution, but they only capture the surface signature of the ocean dynamic and mesoscale eddies. 
On the other hand, \textit{in situ} measurements collected by
autonomous probes such as the ARGO floats~\cite{IEEEhowto:Riser} are optimally providing one vertical profile of temperature (and salinity) measurements over $300km^2$ per month, which is way too sparse to sample mesoscale eddies; but uniquely provides information about the ocean interior. 
Other observation systems exist, such as gliders, but they are not providing measurements systematically and in real time such as the other two mentioned above.
This makes our understanding of these regions and their behavior under the influence of global climate change further limited by our ability to diagnose their 3D thermal structure and temporal evolution.

Latent class regression methods aiming to predict the ocean temperature profile based on several local linear relationship with satellite data (Sea Surface Temperature, SST, and Sea Level Anomaly, SLA) are used to take into account the presence of several dynamic modes observed in turbulent regions~\cite{IEEEhowto:Tandeo}, each mode being characterized by local regression. \citet{IEEEhowto:Maze} introduced an unsupervised classification of oceanic profiles that provides a direct probabilistic link between the vertical oceanic structure of a measurement and the dynamic modes. In this paper, we develop several strategies to combine satellites and \textit{in situ} ocean measurements to produce a 3D time series of temperature with high horizontal resolution.  By leveraging knowledge from the former unsupervised classification of oceanic profiles, we demonstrate that latent class regression methods improve the prediction of vertical ocean temperature fields. 
\begin{figure*}
  \includegraphics[width=\textwidth]{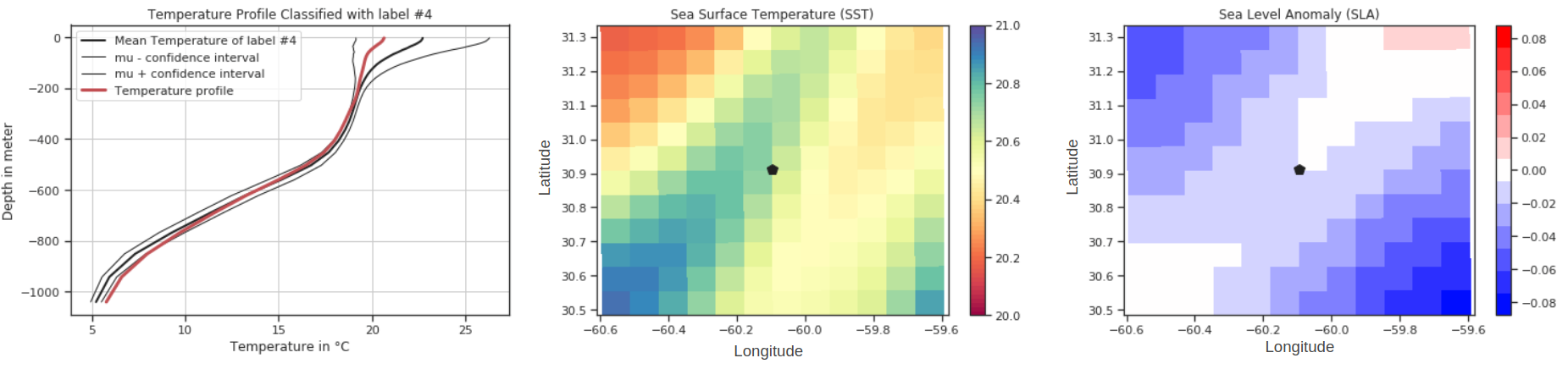}
  \vspace*{-5mm}
  \caption{Overview of the data: On the left a vertical temperature profile between 0 and 1000 meters belonging to the mode \#4. In the middle a co-localised $13\times13$ pixels patch of SST and on the right of SLA. One pixel is about $70km^2$.}
  \label{sstsla}
\end{figure*}

\section{Data}

In a real-case scenario we would be using remote sensing (SST and SLA) and \textit{in-situ} data taken from the ARGO database.
ARGO is a real-time global ocean observation network consisting of about 3000 autonomous profiling probes randomly distributed in all oceans and measuring pressure, temperature and salinity.
The ARGO database is a collection of temperature and salinity vertical profiles going from the surface down to 2000 m, evenly distributed throughout the seasonal cycle and with approximately one profile per month per $3^o$x$3^o$ cell between 2000 and 2014. Routinely used in physical oceanography, these data are key to the observation of the ocean climate \cite{IEEEhowto:Riser}.

However, in order to keep our demonstration away from technical difficulties associated with real-world observations, in this study we will use synthetic observations, i.e. data generated with an ocean general circulation model \cite{drakkar} that provide all necessary data on a 3D grid representing the North Atlantic Ocean, i.e. full data every 8-km. This is a state of the art synthetic set of observations that has been shown to reproduce the ocean structure and dynamic faithfully, especially the turbulent Gulf Stream \citep{maze-2013}.
More precisely we will sub-sampled the synthetic set of observations to mimic the real sampling of ARGO sparse probes.

\begin{figure}[h!]
  \includegraphics[scale = 0.305]{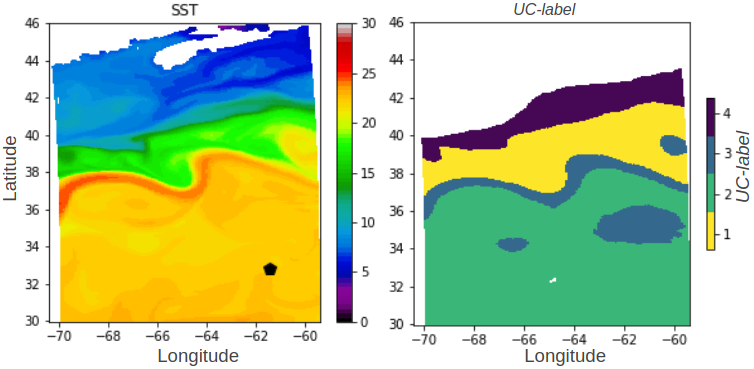}
  \caption{On the left the SST field covering the zone of study. On the right the \textit{UC-label} distribution. Both fields vary with time.}
  \label{uc-label}
\end{figure}

Given Sea Surface Temperature (SST) and Sea Level Anomalies (SLA) patches of size $13\times13$ pixels covering approximately 70km$^2$, we develop a model to predict the associated Vertical Temperature Profile at the center of the patches. A temperature profile represents the temperature measured at 36 different depths distributed between 0 and 1040 meters below the surface. We note $X$ the concatenation of (SST, SLA) and $Y$ a profile of temperature. The dataset of study used for training and evaluation is composed of 700 pairs of vectors $X$ and $Y$ sampled at different locations in space and time inside the region of study, with 66 \% of the data kept to fit the model and the rest used for evaluation. In this document we will denote by \textit{UC-label} the four labels from the unsupervised classification of profiles applied to the this domain and following \citet{IEEEhowto:Maze}. We can clearly see the different water masses with cold and warm temperature at the surface in Figure \ref{uc-label}-left, and matching regions with specific \textit{UC-label}. \textit{UC-label} 1 and 2 denote the cold and warm flanks of the Gulf Stream, while the \textit{UC-label} 3 and 4 denote the frontal regions of the Gulf Stream and continental shelf zones, respectively. 

Figure \ref{sstsla} presents a sample of the data set; the black dot in the center of SST and SLA patches is the location of the vertical temperature profile that is used to produce \textit{UC-label} in \ref{uc-label}-right. The training and evaluation set are voluntarily small to stay true to the amount of ARGO profiles we would have access on a short period of time. The test set is however composed of 1000 randomly chosen profiles and 27 different time-step equally spaced in time over a year period.

\section{Latent Class Regression Model}
Motivated by the fact that one SST/SLA relationship can arise from different vertical temperature profiles but that one vertical temperature profile correspond to a single SST/SLA relationship, our objective is to identify $K$ modes (latent variable $\mathcal{Z}$) from a joint set of SST and SLA patches ($p$-dimensional vector $X$) adapted to predict the temperature profiles. Since \cite{IEEEhowto:Maze} have suggested that a given dynamic region (hence caracterised by a single SST/SLA relashionship) is associated with a unique vertical profile class, we will assume here that \textit{UC-label} are an appropriate choice to learn local linear relationship between SST, SLA and the temperature profile.
We assume that the conditional probability of $Y$ given $X$ and a dynamical mode $\mathcal{Z} = k$ is given by equation \ref{eq:2}.
\begin{equation}
\label{eq:2}
    p(Y|X,\mathcal{Z}=k) \sim \mathcal{N}_k(Y;X \beta_k, \Sigma_k)
\end{equation}
Where $\mathcal{N}_k$ represents a multivariate Gaussian probability density function evaluated in $Y$ with mean $X\beta_k$ and covariance $\Sigma_k$. Hence, the conditional probability of $Y|X$ resorts to a mixture of normal distributions, where $\lambda_k$ is the prior probability of mode $k$. :
\begin{equation}
\label{eq:3}
    p(Y|X,\Theta) = \sum_{k=1}^{K} \lambda_k \mathcal{N}_k(Y;X \beta_k, \Sigma_k)
\end{equation}
To simplify the notations, we store the overall parameters of equation \ref{eq:3} in $\Theta=(\lambda_1,..,\lambda_K,\beta_1,..,\beta_K, \Sigma_1,..,\Sigma_K)$. In the literature this model is referred to as a "latent class regression". The maximum likelihood estimation procedure for model parameters $\Theta$ is given in \citet{DeSarbo1988AML}. A key aspect of the latent class regression is the choice of $K$ the number of latent mode. A popular method to estimate $K$ is the Bayesian Information Criteria \cite{schwarz1978}. The BIC is an empirical approach of the model probability computed as: 
\begin{equation}
   \mathrm{BIC}(K)= -2\mathcal{L}(K) +\mathrm{N}_f(K) \it{\log(n)}
\end{equation}
where $\mathcal{L}(K)$ is the log likelihood of the trained model with $K$ classes. $\mathrm{N}_f(K)$ is the number of independent parameters to be estimated, in our case the parameters to estimate are the one of $\Sigma_k$ the covariance matrix and $\beta_k$  the local linear coefficient. For a given number of mode $K$, we suggest different evaluations of the Expectation Maximization (EM) algorithm \cite{dempster1977maximum}. The idea is to test different initialization of $\hat{\pi}_k$ (probability of belonging to mode $k$) for each sample for the EM procedure and then select parameter estimates corresponding to the greatest likelihood.


The main difficulty in training a latent-class regression method with oceanic profiles is that the dimensions along the patches and the vertical axis are large compared to the number of samples: we have to deal with a vector $Y$ of size 36 and images of $13\times13$ of SST and SLA. This large number of dimensions translates into a large number of parameters to estimate in the covariance matrices $\Sigma_k$ and in the local linear coefficient $\beta_k$. To reduce the number of dimensions, we use Principal Component Analysis. We chose to keep 99.9 \% of variance for both vectors $X$ and $Y$. Once we have projected $X$ and $Y$ in the PCA space, we can train our model with $K$ varying and compute the BICs to set the best probabilistic $K$. From a BIC analysis we learn that the best $K$ is 3. In the following section we compare $K$=3 (best BIC) to $K$=4 motivated by the fact that there are four \textit{UC-label}.





\section{Results and Discussion}
To evaluate our prediction, we observe the Mean Absolute Error (MAE) on the temperature profile. In order to justify every increase in complexity in our approach, we first need to set up a baseline. We keep two different values for the number of modes used for the latent-class regression: $K\in $\{3,4\}. We compare the latent-class regression against a Linear model, a $K$-means initialization of EM algorithm, and a small Convolutional Neural Network. For the CNN approach, we use a small 8-layer architecture with two convolutional layer (32 filters of size $7\times7$) and a max pooling layer (2x2 filter and stride of 2), followed by two convolutional layer (64 filters of size $3\times3$), followed by a the same max pooling layer then by a flatten and a dense layer. 

Table \ref{MAE} summarizes the different results in terms of MAE. The linear model with variance kept above 99.9 \% has good performance but it does not take into account difference between mass of warm water and cold ones (encoded by the modes). Although the CNN does not generalize well to the testing set, we notice from a seasonnal study that its predictions throughout the year are robust to seasonal variations compared to the basic linear model that would present seasonal errors. An entry point to improve CNN performance would be to use wider patches of SST and SLA. The latent-class regression has the best overall MAE although it contains  misclassification errors (attributing a wrong label to a given $X$ before applying the regression results in an error of about 3\% of the MAE).

To fully exploit the potential of latent-regression, we plan to include the temporal aspect of ocean currents in the study: for instance, the profile classes can be tracked in space by applying a Kalman filter procedure with a simple evolution model to identify the class of a profile and an adequate linear coefficient. In that case of filtering method, we could keep the fuzzy value of belonging (probability density score) and work with fuzzy state of each profile.

\begin{table}
\scalebox{0.90}{
\begin{tabular}{llll}
\hline
\rowcolor[HTML]{FFFFFF} 
\textbf{Model\textbackslash Metrics} & \textbf{MAE train} & \textbf{MAE Eval} & \textbf{MAE Test} \\ \hline
\multicolumn{4}{c}{\textbf{PCA variance kept = 99.9 \%}}    \\ \hline
Linear Model                            & 1,96                           & 2,09                          & \textbf{1,72}                 \\
KMEANS K=4                              & \textbf{1,54}                          & \textbf{1,86}                 & 1,81                          \\
KMEANS K=3                               & 1,65                           & 1,96                          & 1,98                          \\
\textit{UC-label}                                 & \textbf{1,47}                  & \textbf{1,87}                 & \textbf{1,69}     \\
\multicolumn{4}{c}{\textbf{No PCA}}    \\ \hline
CNN                                                & \textbf{0,87}                  & \textbf{1,27}                 & 2,02                          
\end{tabular}}
\caption{Results of the different latent class regression model against the baseline. $K$ is the number of modes..}
\label{MAE}
\end{table}

\section{Conclusion}

In this study, we used a Latent Class Regression model that is an unsupervised classification - supervised regression method on synthetic ARGO temperature profiles. Our goal was to explore to what extent this method would lead to a prediction of profile temperature sufficiently precise vertically and with a correct resolution horizontally to follow the eddies. We hence have developed several models to build the foundations of 3D ocean temperature field prediction from coupled surface and in-situ data. Although our results are not immediately applicable (since the filtering step is missing), they are a crucial step in evaluating latent regression methods for studying vertical ocean temperature fields. Finally, our approach is one more step to follow the diffusion and distribution of the heat in the ocean and thus to better understand global climate change.

\newpage
\bibliography{example_paper}
\bibliographystyle{icml2019}

\end{document}